# Understanding Mechanical Characteristics of FeNiCrCoCu HEA in Nanoscale Laser Powder Bed Fusion via Molecular Dynamics

Ishat Raihan Jamil, Ali Muhit Mustaquim, Mahmudul Islam and Mohammad Nasim Hasan[*]

Department of Mechanical Engineering, Bangladesh University of Engineering and Technology

Dhaka-1000, Bangladesh.

[*]Corresponding author *Email address:* nasim@me.buet.ac.bd

**Abstract**

The concept of alloying multiple principal elements at high concentrations has led to the development of High Entropy Alloys (HEA) with exceptional mechanical properties, making them the focus of major recent scientific endeavors. Geometrically complex HEAs with tailored microstructural characteristics can be produced using additive manufacturing technologies such as powder bed fusion (PBF). However, an in-depth study on the effect of process thermal conditions during PBF is required to effectively fabricate HEAs with desirable mechanical characteristics. Thus, in our present molecular dynamic (MD) study we have explored the implication of PBF process thermal conditions on the mechanical characteristics of FeNiCrCoCu HEA by systematically varying laser scan speed from 0.4 Å/ps to 0.1 Å/ps, unidirectional and reversing laser passes from 1 to 4, and laser power from 100 µW to 220 µW. Our investigation suggests that reducing the laser scanning speed up to a critical velocity of 0.2 Å/ps considerably improves mechanical strengths, with further reduction creating severe surface defects. Decreased ultimate tensile strength (UTS) is associated with the annihilation of the bulk sessile dislocations during tensile straining marking an early yield failure. Alternately, the material's strength could be improved by annealing with several unidirectional laser passes over the same target region, resulting in enhanced UTS due to subtler yield points. Increasing laser power aids in ameliorating material density ultimately leading to higher UTS even in non-dislocation-free structures. These findings will assist researchers to understand the underlying effects and optimize process thermal parameters to fabricate superior HEAs utilizing additive manufacturing.

# 1. Introduction

High-entropy alloys (HEAs), such as FeNiCrCoCu, are a novel class of multicomponent alloys with significant structural and functional properties. HEAs are loosely defined as solid solution alloys containing more than five major elements in equal or nearly equal proportions [1]. They are high-performance materials having exceptional qualities for applications involving elevated temperatures [2]. They have a high degree of toughness, resistance to high-temperature softening, high ductility, good fatigue resistance, and excellent wear resistance [3–5]. HEAs are alloys with a mixing entropy greater than or equal to 1.5R, where R is the gas constant. [6].

Previous MD study by Li et al. examined abrasive particle and material wear and damage in order to determine the wear performance and protection ability of a FeNiCrCoCu HEA coating on a Cu substrate [7]. Cai et al. [8], on the other hand, investigated the improvements of Cu addition on the hardness, wear, oxidation resistance of the HEA. Kretova et al. used molecular dynamics methods to determine the properties of the $Fe_{20}Ni_{20}Cr_{20}Co_{20}Cu_{20}$ high-entropy alloy's interstitial atoms and vacancies [9]. Furthermore, Verma et al. explored the influence of alloying elements on the microstructure development and wear characteristics of FeCoCrNiCu HEA [10]. Additionally, Tian et al. [11] investigated the atomistic effects of twin boundaries on the nanoindentation implications of the same alloy.

The demand for and use of 3D printing has increased significantly in recent years as printing equipment has become more inexpensive and user-friendly. Additive Manufacturing (AM) is a type of 3D printing in which an item is constructed layer by layer by depositing materials in accordance with specified digital data [12–14]. Additive Technology has transformed the manufacturing business. Compared to traditional production, AM saves cost, time, and carbon impact. Complex geometries may be fabricated and customized without tools or molds. Using optimal settings reduces porosity and improves mechanical characteristics. As a result, it is extensively employed in the aerospace, automotive, medical, and architectural fields [14–16].

Powder bed fusion (PBF) technology of 3D printing uses a heat source, such as a laser or electron beam, to combine powdered materials into three-dimensional objects according to computer-aided design (CAD) data. As each layer is fused either through complete laser melting or through partial melting by laser sintering, it progressively drops down and new powders are spread over it [17,18]. The finished product has a near-net form with a relative density of up to 99.9 percent [19]. This process is associated with high porosity [20,21] and deformation [22,23]. In the PBF process, melting and solidification occur on very small lengths and time scales. Experimental examination of thermo-fluid-mechanical behavior is thus challenging due to the high cost, limited resolution, and complexity of measurement techniques. During PBF processes, such as the novel micro selective laser melting (μ-SLM) technique, the melting phenomena are governed by nanoparticle composition, something which continuum approaches strive to portray. However, such emerging processes can also be investigated at the nanoscale with the help of molecular dynamics (MD) studies as demonstrated by Kurian et al. [24]. Thus, MD simulations might well be a useful tool for visualizing and understanding the effect of modifying certain process parameters, such as laser scan speed and powder, on a material's overall microstructural and mechanical properties.

Recently, researchers have begun investigating innovative AM techniques aimed at finding a balance between high-speed additive manufacturing, printing precision, and functional material characteristics [25]. Previously conducted MD investigations used laser velocities up to 2.0 Å/ps [26]. In addition to being computationally more beneficial and less time-consuming to simulate at the nanoscale, high-velocity melting of a single layer of powder would eventually provide great understanding of the atomistic impacts of the high-speed additive manufacturing process with newer emerging technologies such as μ-SLM, especially with emerging materials such as FeNiCrCoCu HEA. However additional concerns, to the best of the authors' knowledge, remain unanswered from previous literature: how different laser configurations, such as laser scan speeds, interact with the thermal properties of the material thereby affecting the strength of a HEA such as FeNiCrCoCu, and would multiple laser passes at higher speeds or higher laser power for singular passes be reasonable alternatives to reducing velocity for advances in strength improvements. As a result, in this study, we explored the effects of varying laser configurations on the powder bed fusion of FeNiCrCoCu HEA, as well as the specific contributions of temperature profile and certain dislocations in the alloy's strength properties.

## 2. Methodology

### *2.1 Atomic Structure Details*

The simulation model in this study, as illustrated in fig. 1, consists of 105 solid equimolar FeNiCrCoCu high entropy alloy (HEA) powders residing over a solid iron (Fe) bed. The simulation domain is considered to have a orientation of: X [1 0 0], Y [0 1 0], and Z [0 0 1], with periodic conditions in the X and Y directions. The Z boundary is configured in such a way that it confines the atoms in the vertical direction. The bed, having a dimension of 25 nm × 45 nm × 1.7 nm, is made up of 176400 Fe atoms organized in 10 monolayers. The base two atomic monolayers are held fixed (Z: 0 nm ~ 0.2 nm) while the following two monolayers (Z: 0.2 nm ~ 0.6 nm) act as the thermostat layers. The top six Fe monolayers (Z: 0.6 nm ~ 1.7 nm) function as a conduction medium, transferring thermal energy from the heat source to the thermostat bed layers, which serve as a means of energy dissipation for the system. The equimolar HEA powders are arranged in three rows 35 spheres over the substrate bed. Each spherical powder comprises 5727 atoms and has a radius of 2.49 nm, similar to previous studies [26–30], thus bringing the total height of the powder layer to 14.94 nm, above the substrate. In reality, the powers may be of different sizes, however, representing them in such a uniform manner opens up scope for further studies, facilitating a more realistic interpretation.

## *2.2 Interatomic Potential*

The interatomic interaction forces of the FeNiCrCoCu HEA have been characterized using an embedded atom method (EAM) potential developed by Deluigi et al. [31]. In this method, positively charged metal ions are considered to be immersed inside a localized electron density. Thus both the embedding and ion repulsion energy contributes to the overall energy of the system, which is approximated by:

$$E_{total} = \sum_i \left[ F_i(\rho_i) + \frac{1}{2} \sum_{j(\neq i)} \Phi(R_{ij}) \right] \qquad (1)$$

where $\Phi(R_{ij})$ is the potential of pair interaction between atoms $i$ and $j$ having a separation of $R_{ij}$ [32]. The embedding function $F_i(\rho_i)$ represents the energy needed to locate the atom $i$ in the host electron density $\rho_i$, which the EAM method determines by superimposing the atomic densities of the neighboring atoms at atom $i$:

$$\rho_i = \sum_{j(\neq i)} \rho_j^a(R_{ij}) \qquad (2)$$

where $\rho_j$ is the electron density on another atom $j$ isolated from the nucleus of atom $i$ by a distance of $R_{ij}$ [33].

## 2.3 Simulation Procedures

The motion of the powder layer, as well as the upper eight Fe monolayers of the bed, is integrated using the Velocity-Verlet algorithm. A Langevin thermostat is used to set the temperature of the thermoset bed layer atoms to 300 K. A time step of 1 fs is used for the computational calculations of the simulation domain. Initially, the powders are laid over the iron bed, and the whole simulation cell is equilibrated for 400 ps using the microcanonical ensemble (NVE). Prior literature dictates two methods for simulating PBF at the nanoscale: either by modeling a laser in finite element analysis (FEA) method and applying the resulting temperature profile in a MD simulation domain containing powders using a thermostat [24], or directly implementing the laser in MD domain by modeling a dynamic region that would add energy to the system [26,28–30]. The former method is limited by the accuracies of both FEA and MD simulation, whereas the latter method presents opportunities for varying several laser parameters such as laser scan speed, passes, and power very easily. As such, a cylindrical heating region with a diameter of 10 nm [27,29,30] is modeled to continuously supply energy to the system and melt the HEA powders as it travels along the Y-axis, as shown in fig. 1. The rate of motion of this dynamic region within the simulation domain is determined by the laser scan speed selected. Former MD studies have used scan speeds in excess of 0.5 Å/ps [26,27,30], while laser power in ranges of 80-128 µW had been implemented by Zhang et al. [30]. This range of laser powers also corresponds with the max power density of $2.8 \times 10^{12}$ W/m$^2$ in some current 3D laser sintering systems [34], which translates to about a maximum power of 222 µW at the nanoscale. Following the completion of the laser movements, the laser is turned off the whole simulation cell is allowed to cool down to 300 K within 2000 ps.

Once the system is cooled, a representative block of dimension 8 nm × 20 nm × 5 nm is cut away from the middle of the resolidified FeNiCrCoCu HEA by discarding loose atoms in the vicinity, following Kurian's [24] approach as also illustrated in fig. 1. This section is common in all the simulations and disregards the bottommost row of powders, which usually remains partially unmelted. As this segment is isolated from the main simulation domain, shrink-wrapped conditions are implemented at the boundaries. Thus it requires the use of the method demonstrated by Kurian et al. [24] to perform the tensile and compression test on the isolated blocks. The block is initially relaxed utilizing the conjugate gradient approach for energy minimization. With a time step of 0.5 fs, it is then equilibrated for 60 ps at 300 K and 0 bar pressure in the Y-direction, using an NPT ensemble. The ensemble implements non-Hamiltonian equations of motion in the Nose-Hoover style to carry out time integration. Following equilibration, two thin layers of atoms at the opposing ends of the block along the Y-direction are displaced either towards or away from each other at a constant engineering strain rate of $10^9$ s$^{-1}$ to perform the strain-controlled uniaxial compression or tensile test, respectively, as dictated by Kurian et al. [24]. Other MD studies have also used such high strain rates [35–40], and according to Zhang et. al. [37] notable difference in the material strength is only observed past $10^9$ s$^{-1}$ strain rate. Thus, to save computational time, the block is strained up to $\varepsilon$ = 20% of the original Y dimensional length, at this rate, in either direction in separate simulations as shown in fig. 1.

The large-scale atomic/molecular massively parallel simulator (LAMMPS) [41] is used to perform all the molecular dynamics simulations of this study. The data acquired from these simulations are subjected to post-processing analysis, such as common neighbor analysis (CNA) [42] and dislocation analysis (DLA) [43], using the open visualization tool Ovito [44].

## 3. Results and Discussions

### 3.1 Process and Performance Parameters

Numerous simulations have been performed to determine the atomistic effects of various scan speeds, effects of multiple laser passes and the influence of increasing laser powers on the mechanical tensile and compressive strength ($\sigma$) of the resultant block. To ensure comparability, the bed temperature is maintained at 300 K throughout the investigations. Table 1 summarizes the details of the other parameters. The laser is always moved in a single direction for all cases apart from H, I, and J multi-pass simulations in which the direction of laser travel is reversed after each pass, as illustrated in fig. 2. Case A is regarded as the baseline simulation since it is applicable to all parametric investigations discussed in the subsequent sections. The table also list the laser energy densities ($E_d$) computed using the equation presented by Donik et al. [45]:

$$E_d = \frac{P}{v \times d \times l} \tag{3}$$

where, $P$ = laser power, $v$ = laser scan speed, $d$ = laser spot diameter, and $l$ = layer thickness.

For a single pass, the motion of the laser beam is depicted in fig. 3. The common neighbor analysis (CNA) of the section view gives a clear understanding of the melting and resolidification mechanism that occurs as the laser moves rightwards in the positive Y-direction. The dynamic heating region continuously adds energy to the system, raising the temperature of the FeNiCrCoCu HEA powders. The powders melt when the temperature exceeds 1400 K. Generally, the atoms of the HEA crystallize in a stable face-centered cubic (FCC) configuration. As the laser passes by during the melting phase, a melting front develops that advance rightwards and downwards at a 45° angle with the positive Y-axis, as seen in fig. 3. This causes the powders to lose their regular forms and the atoms become amorphously dispersed, forming the liquid melt. As the laser advances rightward, more free atoms from progressively melting powders coalesce with the melt pool. As the temperature rises beyond 1400 K the viscosity of the melt reduces and it rapidly fills any unmelted powder spacings. The temperature drops behind the laser as thermal energy is conducted downward and dissipated by the thermostat bed layer atoms. This leads to the formation of a solidifying front that proceeds rightward and upward at a 65° angle to the positive Y-axis, with its velocity dictating the cooling rate. During the cooling phase, the atoms in the melt pool try to settle into a stable FCC lattice. However, the resolidification process is far from perfect owing to the haphazard distribution of the atoms in the liquified alloy. The resulting structure contains numerous vacancies, dislocations, and stacking faults as visible from the CNA and DLA in fig. 3. The vacancies can be described as void regions within the FCC matrix where an atom or a group of atoms are missing, which is an indication of the occurrence of rapid cooling. Qi et al. previously demonstrated the effects of different sized voids on a similar HEA [46]. However, in the present study the vacancies, although numerous, are much smaller in size and thus require further investigations. When a representative block is extracted from this solidified structure, the dislocations contained within it are quantified using the dislocation length $L_d$. The stacking faults, as can be seen in fig. 3, are essentially atoms settling in hexagonal close packing (HCP) between two partial dislocations. The fraction of atoms in the HCP stacking defects is expressed in similar manner to %FCC as follows:

$$\%HCP = \frac{nHCP}{N_{atoms}} \times 100\% \tag{4}$$

where $nHCP$ denotes the number of atoms in the HCP lattice inside the representative block and $N_{atoms}$ denotes the total number of atoms contained within the same block.

*3.2 Effect of Laser Scan Speed*

To investigate the effect of laser scan speed, the laser power (P) and number of laser passes (N) are kept constant at 100 µW and 1, respectively. According to equation 3, this means that $E_d$ only varies inversely to the scan speed. The sectional CNA during the melting phase for N = 1 at laser scan speed (S) of 0.4 Å/ps is shown in fig. 3, while the CNA of the post-cooled structure and the DLA of the cut-off representative blocks are displayed in fig. 3. The figure also shows the post relaxed blocks which results from the energy minimization and NPT equilibration step that alleviates some of the internal stresses at the start of the tensile and compression processes, leading to the removal some of the dislocations and stacking faults from the blocks. Fig. 5(a) illustrates the temperature profile of a central 10 nm cylindrical region in the simulation domain during the PBF process, while table 2 summarizes its thermal characteristics. Fig. 5(a) shows that there are two distinct heating and cooling rate above and below 700K. The rates calculated in the table are of those above this temperature when the laser passes by the observed region, since those below are merely preheating of the same region due to the far field effect of laser heating. The stress-strain curves of the representative blocks obtained from the uniaxial tensile and compression test are shown in fig. 5(b) and fig 5(c), respectively. Despite the strengths appearing to be higher in the figures than what is to be anticipated, previous MD studies, such as by Faiyad et al. [47], validates it by reporting strengths in similar ranges. This is a likely result of determining strengths at the nanoscale where fewer defects, such as vacancies, are expected to exist compared to the macroscale. Approximation in the EAM potential, which defines the interactions between the atoms, also contributes to this divergence.

Slowing down the laser scan speed significantly improves the tensile yields strength (TYS) and the ultimate tensile yield strength (UTS) of the manufactured block, but up to a certain limit. For S = 0.4 Å/ps or $E_d$ = 14.34 J/mm$^3$, fig. 5(b) reveals that the block achieves a UTS of 6.58 GPa, while TYS occurs at nearly half of that value owing to the combination of a fast-moving melting front with a very high melt temperature ($T_{max}$) and a subsequent fast cooling rate, as understandable from fig. 5(a) and table 2. The CNA in fig. 3 and fig. 4 shows that such circumstances result in fair number of dislocations and HCP stacking faults in the FCC lattice. The remaining atoms in the figures, mostly those around the vacancies, are identified as amorphous atoms that eventually influence the material density (ρ) of the block. Raising $E_d$ to 19.12 J/mm$^3$ by reducing the S to 0.3 Å/ps slows down the melt front heating but $T_{max}$ attained remaining virtually unchanged as before.

The cooling rate is also now visibly slower. Compared to S = 0.4 Å/ps, a high $T_{max}$ coupled with a slower front heating and cooling rates results in an increase in the amount dislocations as well as the %HCP, as also visible fig. 4. Interestingly, this does not seem to have negative effects on the strengths, rather improving the TYS and UTS by 91.8% and 5.01 %, respectively. The peak strength of 7.97 GPa is obtained for S = 0.2 Å/ps or $E_d$ = 28.69 J/mm$^3$, at about ε = 7.5%. At this speed, both TYS and UTS occur at the same point. A comparably low $T_{max}$ of about 1925 K in conjunction with a smaller difference between the heating and cooling rates appears to be just sufficient enough to induce the least dislocations and %HCP stacking faults as also evident from fig. 4. Decreasing the S to 0.1 Å/ps does not show any improvement, even though $E_d$ is further increased, but rather decreases the TYS and UTS by 11.7% and 9.54 %, respectively. At such comparatively low speed, the melt finds a sufficiently long period to disperse the thermal energy away from it, thereby attaining a low $T_{max}$ of only 1872 K and an even smaller difference between the fronts' heating and cooling rates. Thus, when the upper melt atoms move outwards and away from the heat source as the laser passes by, they solidify more quickly than are unable to reflow back, therefore leaving depressions on the laser track. Infliction of such severe surface defects not only weakens the block but also lowers the ρ to 7.57 g/cm$^3$ from about 7.8 g/cm$^3$, as was the case for the prior speeds. Nonetheless, in all instances, the stiffness of the material remains roughly unchanged at about 99.74 GPa.

For the case of the uniaxial compression test, the results are a bit more complicated. During the test the atoms are forced onto each other, resulting in a complex interaction between the dislocations and stacking faults thus translating into certain waviness of the compressive stress-stain curves. The presence of such fluctuations in the curves, along with the closeness of the compressive yield strengths (CYS) and the ultimate compressive strengths (UCS), makes it difficult to distinguish them from one another in many cases. As it appears from fig. 5(b), UCS for both S = 0.4 Å/ps and S = 0.3 Å/ps occurs at about 3.6 GPa, but at different strains. Interestingly, the greatest UCS of about 4.7 GPa is obtained for both S = 0.2 Å/ps and S = 0.1 Å/ps, implying that variation of laser speeds has little effect on the compressive strengths of the manufactured blocks.

Thus, it appears that the temperature profile during the PBF process play important role in determining the strengths of the manufactured HEA alloy blocks. Both UTS and UCS appear to be at maximum values for the conditions inflicting the least dislocations and stacking faults, such as for S = 0.2 Å/ps when $E_d$ = 28.69 J/mm$^3$. On the other hand, when substantial proportions of dislocations exist as is the case for the other speeds, with the exception of S = 0.1 Å/ps, the dislocation lengths seem to be correlated with the strengths when the material density virtually remains unchanged, analogous to the effects of work-hardening processes [48]. Interestingly, Joo et al. noted that HEA exhibits higher strain-hardening than low and medium entropy alloys [49]. The specific significance, however, of each type of dislocation in the yield and ultimate strengths of the material is obscure from the surface level. As such, further analysis on them is carried out in section 3.4.

*3.3 Effect of Multiple Laser Passes*

Instead of reducing the laser scan speed, passing the laser multiple times over the target area at higher speeds might be yet another way of improving the microstructure and eventually the strength of the manufactured product. To investigate this hypothesis, three more simulations are performed on top of the baseline simulation A. In each case, the laser power, and scan speed are kept constant at 100 µW, and 0.4 Å/ps, respectively. As such, $E_d$ remains virtually unchanged for these simulations. Fig. 6 depicts the sectional view of the CNA of the simulation domain near the end of each unidirectional laser pass, as well as the DLA of the cut-off representative blocks from these domains. The temperature profile of a central region during PBF process and, the stress-strain curves obtained from the uniaxial tensile and compression test of the cut-off blocks are shown in fig. 7. As it appears from the figure, reheating a solidified zone just after a single laser pass can drastically improve the strength of the material block produced. The $T_{max}$ attained in the second pass only reaches about 1515 K, as observed from table 3 and fig. 7(a), marginally above the melting point of the material. Such circumstances influence only partial melting of the solidified alloy as visible in fig. 6. The figure also depicts that sessile, or immovable, dislocations such as Stair-rod and Hirth lock, as well as subsequently connected glissile, or movable, dislocation such as Shockley, retain their position even after successive laser passes. On the other hand, a reduction of viscosity during the remelting process allows the free Shockley dislocations to move downwards away from the heat source during each pass. In other words, such reheating process archives a similar effect to annealing [50], and thus can be regarded as a variation of the heat treatment process. Similar to the preceding section, relaxing the blocks prior to straining them removes most of the glissile dislocations. The greatest difference in the post cooled representative block is observed between N =1 and N= 2 passes, in which a large chunk of an HCP stacking fault plane is replaced by smaller stacking fault planes as some of the free Shockley dislocation lines splits. This is likely due to the influence of a slow but nearly equal heating and cooling rates in conjunction with the low remelt temperature. As a result, UTS increases from 6.58 GPa to 7.44 GPa, being only 6.65% weaker than that obtained with a single laser pass at 0.2 Å/ps. Further passes neither influence the remelt temperature nor effect the front velocities noticeably, as illustrated in fig. 7(a) and table 3. Except for N = 3, the overall effect of 4 passes is a slight 6.45% increase in the UTS of the material block to 7.92 GPa. The reasoning for such ambiguity in otherwise predictable trends in changes of UTS with additional laser passes is further discussed in section 3.4. The annealing process achieved by multiple unidirectional laser passes tends to migrate some of the vacancies away from the central region but does not hamper their count appreciably, leading to $\rho$ remaining unhampered at 7.9 g/cm$^3$. The stiffness also remains fairly unaffected at about 98.58 GPa. In each of the remelt passes YTS occurs at about $\varepsilon = 4.5\%$ but is barely noticeable in the stress-strain graph. On the other hand, the fluctuations in the compression stress-strain curves seem to become more prominent making YCS less discernible in fig. 7(c). UCS increases to about 5.1 GPa for N =2, but unlike the single-pass cases, it mildly improves with increasing N. The experimental observations of a similar HEA to the present alloy by Wang et al. [51] dictates that an optimum annealing temperature of about 900 K and proper control of annealing time is key to achieving improved strength. Although such fine control of these parameters in the PBF process is difficult, the short bursts of moderately high temperature of 1500 K during the remelting process of FeNiCrCrCu HEA appears to be just enough to achieve similar overall effects as the experimentations.

Fig. 7 depicts the tensile and compressions test results of another set of multipass laser simulations with the same parameters and conditions as before, but with the direction of the laser motion reversed after each pass. With essentially the same remelt temperatures, heating, and cooling front speeds as before but with lower overall material density of 7.7 g/cm$^3$, the figures show no added benefits of such configuration, but rather a lower maximum UTS of 7.18 GPa at the end of 4 reversing passes. Again, TYS and UTS seem to occur at the same point. CYS still tends to remain indistinguishable with several fluctuations of the compressive stress-strain curve as visible from fig. 8(b). However, unlike unidirectional laser passes, the UCS does appear to drop off after 4 reversing laser passes.

### *3.4 Role of Dislocations on Mechanical Characteristics*

Previous MD studies focused on tensile tests, at different conditions, starting with either defect-free structures [52] or manually inserted defects [46,47]. The novelty of the present study is that the structures used for the mechanical tests are obtained from another process, that is the PBF. Thus, these structures inherit their defects such as vacancies, dislocations, and stacking faults, making them a better representation of real-world scenarios. Previous discussions in section 3.2 highlighted the implications of the presence and absence of dislocations as a whole, while section 3.3 emphasized the advantages of annealing realized using multiple laser passes on the strength of the FeNiCrCoCu HEA block manufactured using the powder bed fusion process. However, the implication of each type of dislocations on the material strengths are yet to be analyzed. Dislocation lines are simply partitions between slip and no-slip regions within a material. The DLA of blocks in fig. 4 and fig. 6 depict that the stacking faults generally form between a pair of partial dislocations called the $\frac{a}{2}<112>$ Shockley partials, where a is the lattice constant of the HEA. These dislocations are glissile, implying that they are free to glide on the slip panes, {111} in the case of FCC materials such as the HEA in this study. However, when two such glissile Shockley partials on intersecting slip planes, and parallel to the line of intersection, meet at the junction they form a different kind of dislocation known as the sessile $\frac{a}{6}<110>$ stair-rod dislocation, which has lower dislocation energy than its prior constituents. Likewise, when two perfect dislocations on separate but bisecting planes join at the convergence line, a sessile Hirth lock is formed if their individual burgers vectors add up to generate a vector of the form $\frac{1}{3}<100>$ [53]. Sessile dislocations are the opposite of glissile dislocations, meaning that they are unable to glide and remain fixed in place. In a system of several such dislocations, one line could possibly interfere with other dislocation lines, thereby making their motion more difficult and eventually resulting in the strengthening of the material as evident from the findings of section 3.2. This is supported by the observations of both Li et al. [54] and Fan et al. [48] that massive lengths of Hirth and stair-rod dislocations are indicators of the formation of Lomere-Cottrel Lock, which considerably enhances the material's strength. The dislocation lines do not end randomly, but rather terminate on other dislocations, as visible in fig. 3 and fig. 5, at certain points called nodes. Since the blocks are cut away from another simulation domain, some of the stacking fault planes continue to the edges of the blocks as the other half of the dislocation lines required to terminate them are left on the discarded portions of the prior simulation cells.

Alterations in glissile and sessile dislocation lengths, as well as %HCP stacking faults of the blocks under various uniaxial tensile straining conditions discussed previously are illustrated in fig. 9, whereas fig. 10 portrays some of their internal changes. The block manufactured from N = 1 laser pass at S = 0.4 Å/ps achieves TYS at about $\varepsilon$ = 3%. It is associated with a steep but small rise in glissile dislocation length and an even smaller change in sessile dislocation length, as comprehensible from fig. 9(a). Fig. 10 depicts that while straining beyond this point certain lengths of sessile stair-rod dislocation are annihilated and replaced by several segments of glissile Shockley partials and some lengths of Hirth lock dislocations, while the HCP stacking faults remain virtually unchanged. Further staining up to $\varepsilon$ = 5% results in shortening of the glissile dislocations as well as the complete elimination of the sessile dislocations. Beyond this point, the block again shows a linear stress-strain relationship up to the point of failure at about $\varepsilon$ = 7.5% in fig. 5(b). Staining past the UTS point prompts a complete disarrangement of the stacking fault planes and a very sharp rise in the glissile and, a comparatively small increase in sessile, dislocation lengths. Interestingly, the favorable conditions during the manufacture of the block at N = 1 and S = 0.2 Å/ps, means that it is able to start the tensile testing process with minimal dislocations and near-zero %HCP, as can be seen in fig. 9(b). Although there is a slight increase in both Shockley partials and %HCP beyond $\varepsilon$ = 4%, it fairly goes unnoticed in the tensile stress-strain graph. Sessile dislocations remain absent from the block throughout the test until the material eventually fails in a similar manner as before. A similar situation arises for the block obtained from N = 1 at S = 0.1 Å/ps, thus implying that lower tensile yield strength point in the FeNiCrCoCu HEA is only realized if the material has inherent sessile dislocations that eventually break down or are somewhat mitigated during tensile straining. Premature tensile yield failure ultimately contributes to the material's lower UTS. Fig. 10 and fig. 9(c) shows that annealing the material with a second (N = 2) unidirectional laser pass reduces both glissile dislocation length and %HCP substantially before starting the straining process, as compared to the N = 1 case in fig. 9(a), while the sessile length remains unhindered. As such, a hardly distinguishable yield point in the stress-strain graph in fig. 7(b) is prompted by a sharp but small rise in glissile length and a drop in sessile length past 4.5% tensile strain as compared to the non-annealed block produced at the same speed. With both blocks having nearly the same total dislocation length at $\varepsilon$ = 8.0%, the higher UTS of the annealed block could be attributed either to its lower %HCP stacking faults or a substantial presence of sessile dislocations near the point of failure. Fortunately, this query can be resolved by taking a closer look at the case with N = 3. Starting with equal lengths of both types of dislocations, as visible in fig. 9(d), all dislocations and stacking faults get annihilated past 4.0% tensile strength with an even more subtle yield point in the tensile stress-strain graph. However, by $\varepsilon = 7.5\%$, %HCP raises considerably without the presence of any sessile dislocations. The overall effect is a decrease in UTS when compared to a block obtained for N = 2. Hence it can be understood that the presence of stacking faults has little effect, but it's rather the sessile dislocations that play a major role in contributing to the ultimate failure of the material during tensile testing. Unlike single pass case, for multi-pass scenarios such as for N = 2 in fig. 10, twin nucleation becomes prominent with the onset of plastic deformation past UTS point that eventually grows with further elongation.

The changes in dislocation lengths and %HCP during some of the compression tests are shown in fig. 11. For N =1 at S = 0.4 Å/ps, CYS appears at about 3.0% compression strain, with very subtle changes in dislocations but a significant increase in %HCP, as seen in fig. 11(a). From fig. 12 it can be seen the stacking faults mostly appears near the ends along which the block is being compressed. UCS occurs at about $\varepsilon = 4.5\%$ compressive strain, with sessile dislocations being destroyed and replaced by glissile dislocations. For a near dislocation-free structure, the block manufactured with N = 1 and S = 0.2 Å/ps appears to show small variation in glissile dislocation lengths up to $\varepsilon = 3.0\%$ in fig. 11(b), associated with minute fluctuations in the compressive stress-stain in fig. 5(c). A more pronounced plateau is visible in the stress-strain graph at $\varepsilon = 3.5\%$, accompanied by a significant rise in both Shockley partial lengths and %HCP stacking faults. However, sessile dislocations do not emerge until the peak in the stress-strain curve at about 4.0% compressive strain, with a notable increase past the UCS point at about $\varepsilon = 5.0\%$, as observed in fig. 11(b). A similar situation arises for the block obtained N = 1 at S = 0.1 Å/ps, implying that the highest UCS is attainable in the presence of the least number of sessile dislocations. For the case of annealed block produced using N = 2 at S = 0.4 Å/ps, fluctuations in dislocation lengths are observed in fig. 11(c). It is associated with several oscillations in the stress-strain graphs in fig. 7(c), but the most significant change occurs after $\varepsilon = 1.5\%$ with a sudden rise in glissile length as sessile length shortens, making a more distinguishable first peak. Further compression eventually eliminates most of the sessile dislocations. Past 3.0% strain the block remains nearly stacking fault free as seen in fig. 12 until it crosses UCS point at about $\varepsilon = 5.5\%$ strain which results in a sharp rise in %HCP stacking faults and Shockley partials. The block fabricated using a second annealing laser pass (N = 3) shows a more or less similar trend in fig. 11(d) to that treated with a single annealing pass (N = 2), with the exception that all dislocations and stacking faults area annihilated past 3.0% compressive strain. This eventually leads to a higher UCS at about ε = 5.5%. A similar case also occurs in the block produced with N = 4, thus confirming that dislocations and stacking fault free structures achieved during compression testing ultimately contribute to higher UCS. CYS is often indistinguishable attributed to dislocation length fluctuations during compression. However, when the bulk of sessile lengths is annihilated, a considerable decline in compressive stress is observed in stress-strain graphs. Similar to tensile test, compressive straining also reveals the occurrence of twinning past UCS point in fig. 12.

### *3.5 Influence of Laser Power*

The other aspect of laser configuration is laser power. In the preceding sections, the laser power was kept constant at P = 100 µW. However, to further investigate its specific effects the P is increased from 100 µW to 220 µW in compliance with previous MD studies [30] and commercially available 3D metal sintering systems [34], while moving the laser at a constant speed of S = 0.4 Å/ps, for N = 1. Thus, $E_d$ is varied proportional to P. Fig. 13 shows the laser powder melting process at various laser powers. As it appears in the figure, the amorphous melt pool extends as a semi spherical shaped region around the dynamically moving laser, as was also observed by Kurian et al. [24], with its diameter being a function of its input power. In prior cases, the $T_{max}$ only reached as high as 1980 K for P = 100 µW at S = 0.4 Å/ps, with slight reduction at slower velocities. As depicted in Fig. 14(a) as well as tabulated in Table 4, it is evident that increasing laser power, and hence $E_d$, while moving it at a constant speed progressively rises $T_{max}$ of the melt pool as well as the melting front heating rates. This all results in the enlargement of the melt pool diameter, extending it beyond the laser heating zone at any given time, as seen in fig. 13. The solidification front cooling rates behind the laser are also sped up, owing to proper melting and hence better heat dissipation at the bottommost powder layer. Therefore, the resulting representative blocks obtained after cooling shows high proportions of glissile dislocations as compared to sessile dislocations for higher melt temperatures, most of which are annihilated following the relaxation step for removing internal stresses prior to straining them.

Figures 14(b) and 14(c) depicts effects of laser power on tensile and compressive straining of representative internally stress relived blocks. As shown in Fig. 14(b), increasing laser power improves the overall UTS of the alloy blocks of the PBF process. Increasing $E_d$ by 40% by raising P from 100 µW to 140 µW improves the UTS by 12.5%, in the absence of a premature yield point. For the case of P = 140 µW and P = 180 µW, both representative relaxed blocks are considered dislocation-free as the atomic layers near the extreme Y edges, and hence the dislocations around them, are frozen for the process of straining them. Since the additional rise in laser power yields no improvement in UTS, the difference in the vacancy concentration, as suggested by the variance in their material densities from $\rho$ = 7.63 g/cm$^3$ for P = 140 µW to $\rho$ = 7.88 g/cm$^3$ for P = 180 µW laser-treated blocks, likely have very little contribution to material strengths as compared to dislocations. For compressive staining test, increasing laser power from 100 µW to 140 µW improves the UCS by 48.8% to 5.40 GPa, while further rise in laser power lowers it. as illustrated in fig. 14(c). Meanwhile, CYS remains indistinguishable.

During tensile straining, dislocations and stacking faults does not appear for P = 140 µW laser treated block until 8% strain at the time of tensile failure as visible in fig. 15(a). A Similar occurrence also transpires while tensile straining P = 180 µW laser treated block. Intensifying the power to 220 µW or $E_d$ to 31.55 J/mm$^3$, however, shows mild improvement in UTS to 7.9 GPa at about $\varepsilon$ = 8.3% due the amelioration of $\rho$ to 8.26 g/cm$^3$ along with the retention of some sessile dislocations in the final block which remains mostly unchanged during the straining process as visible in fig. 15(b), thereby making the motion of the atoms more difficult. Increase in %HCP and glissile dislocations are only observed for $\varepsilon$ > 7.5%, a common indication of tensile failures as also observed in the proceeding section. Conversely, this contradicts the findings of section 3.2 in which it was concluded an absence of dislocations at certain laser scan speeds yields a higher UTS compared to speeds that induce dislocations within the block region. In reality, the fact that the $\rho$ virtually remains unchanged for reductions in laser scan speeds as opposed to improving with laser power, helps to explain such disparity. However, the average stiffness of the blocks remains nearly unaltered at about 101.86 GPa.

During compressive staining of P = 140 µW laser treated block, no major changes in dislocations and stacking fault are visible in fig. 15(d) until UCS point at $\varepsilon$ = 5.5%, even though fluctuations are seen in compressive stress-strain curve. Additional increase in $E_d$ and hence P to 180 µW slightly lowers the UCS by 8.5%. Raising the laser power any further does not improve the UCS appreciably. Fig. 15(d) shows that there are drop in glissile dislocations and %HCP at about $\varepsilon = 1.5\%$ leading to a plateau and distinguishable CYS for P = 220 µW block, and a major rise in all dislocations for $\varepsilon > 5.0\%$. Thus both fig. 15(c) and fig. 15(d) supports the observation in the previous section that the highest UCS is obtained for the least dislocations and stacking faults.

## 4. Conclusion

The present molecular dynamic study explored the thermal implications of different laser configurations, such as scan speed, passes, and power, on the formation of dislocations in FeNiCrCoCu HEA during the powder bed fusion process. Additional atomistic insights demonstrated how these various types of dislocations eventually affect the mechanical characteristics of the manufactured representative alloy product tested with a strain rate of $10^9$ s$^{-1}$. The study's findings can be summarized as follows:

- Slowing down the laser up to a critical scan speed of 0.2 Å/ps or $E_d$ of 28.69 J/mm$^3$ during the powder bed fusion process improves both TYS and UTS to 7.97 GPa owing to favorable melt pool temperature, heating, and cooling rates. While UCS attains a lower value of 4.7 GPa but CYS remains fairly indistinguishable as was the case for the other conditions. Beyond the optimal speed, changing the laser energy density by increasing or reducing the laser velocity leads to an overall decrease in the strengths. For similar material densities, higher material strengths prefer a dislocation free structure.
- Alternative to lowering scan speed and changing $E_d$, passing the laser multiple times over the target area in the same direction also improves the TYS, UTS, and UCS drastically, attaining a similar effect to annealing. With just two unidirectional laser passes, it is possible to obtain a UTS of 7.44 GPa and a UCS of 5.1 GPa, very close to that obtained of a single laser pass at the optimal speed. Reversing the direction of the laser after each high-speed pass does not result in any substantial improvements, but rather produces UTS and UCS that are either similar to or worse than those obtained for an equivalent number of unidirectional laser passes.
- Increasing laser power helps the melt pool to raise $E_d$ and achieve faster heating and cooling rates due to combination of higher melt temperature and improved heat dissipations. This results in the amelioration of material density with power ultimately leading to the material

- achieving a higher UTS, of 7.9 GPa for 220 µW or 31.55 J/mm$^3$, even with a non-dislocation free structure.
- Dislocation analysis reveals that for a single laser pass, tensile yield strength in the FeNiCrCoCu HEA is only realized if the material has inherent sessile dislocations that eventually break down or are attenuated during tensile straining. Such premature tensile yield failure ultimately contributes to the material's lower UTS. Annealing the material with multiple laser passes causes the fracture of sessile dislocations to put a subtle impression on the tensile stress-strain curves, thus helping to achieve higher UTS.
- Dislocation analysis further demonstrates that the maximum UCS is attainable in the presence of the least number of sessile dislocations and stacking faults. Frequent fluctuations in dislocation lengths often make CYS difficult to discern during the compression test. When the majority of sessile lengths are eliminated, however, stress-strain graphs demonstrate a significant reduction in compressive stress.

Conclusions of the present study are in context to the nanoscopic scale of the simulation domain. In future, microscopic simulations and experimentations will be developed to relate these findings with macro scale. We hope present computational study will help the AM community to better understand the effects of laser configuration on PBF process thermal condition and associated nucleation-growth of dislocations as well as subsequent evolution of mechanical properties of FeNiCrCoCu HEA.

**Declaration of competing interest**

The authors declare that they have no known competing financial interests or personal relationships that could have appeared to influence the work reported in this paper.

**Acknowledgement**

The authors gratefully acknowledge the high-performance computing facilities provided by the Institute of Information and Communication Technology (IICT), BUET in conducting this research. The authors appreciate thoughtful discussion in meeting with Multiscale Mechanical Modeling and Research Networks (MMMRN).

| List of Table Captions | |
|---|---|
| **Table 1** | Details of variations of process parameters for each simulation |
| **Table 2** | Thermal characteristics of PBF process at different laser scan speeds |
| **Table 3** | Thermal characteristics of PBF process at different unidirectional laser passes |
| **Table 4** | Thermal characteristics of PBF process for different laser powers. |

**Table 1**. Details of variations of process parameters for each simulation.

| Case | Process Parameters | | | | Laser Energy Density, $E_d$ (J/mm$^3$) |
|---|---|---|---|---|---|
| | Scan Speed, S (Å/ps) | Number of Laser Passes, N | Direction of Laser Passes | Laser Power, P (μW) | |
| A | 0.4 | 1 | Unidirectional | 100 | 14.34 |
| B | 0.3 | 1 | Unidirectional | 100 | 19.12 |
| C | 0.2 | 1 | Unidirectional | 100 | 28.69 |
| D | 0.1 | 1 | Unidirectional | 100 | 57.37 |
| E | 0.4 | 2 | Unidirectional | 100 | 14.34 |
| F | 0.4 | 3 | Unidirectional | 100 | 14.34 |
| G | 0.4 | 4 | Unidirectional | 100 | 14.34 |
| H | 0.4 | 2 | Reversing | 100 | 14.34 |
| I | 0.4 | 3 | Reversing | 100 | 14.34 |
| J | 0.4 | 4 | Reversing | 100 | 14.34 |
| K | 0.4 | 1 | Unidirectional | 140 | 20.08 |
| L | 0.4 | 1 | Unidirectional | 180 | 25.82 |
| M | 0.4 | 1 | Unidirectional | 220 | 31.55 |

Table 2. Thermal characteristics of PBF process at different laser scan speeds.

| Case | Scan Speed, S (Å/ps) | Melt Front Heating Rate (K/ps) | Solidifying Front Cooling Rate (K/ps) | Maximum Melt-Pool Temperature, $T_{max}$ (K) |
|---|---|---|---|---|
| A | 0.4 | 7.86 | 4.83 | 1980 |
| B | 0.3 | 5.93 | 3.68 | 1979 |
| C | 0.2 | 3.44 | 2.44 | 1925 |
| D | 0.1 | 1.54 | 1.19 | 1872 |

Table 3. Thermal characteristics of PBF process at different unidirectional laser passes.

| Case | Number of Laser Passes, N | Melt Front Heating Rate (K/ps) | Solidifying Front Cooling Rate (K/ps) | Maximum Melt-Pool Temperature, $T_{max}$ (K) |
|---|---|---|---|---|
| A | 1 | 7.86 | 4.83 | 1980 |
| E | 2 | 3.80 | 3.57 | 1515 |
| F | 3 | 3.84 | 3.35 | 1471 |
| G | 4 | 3.81 | 3.39 | 1484 |

Table 4. Thermal characteristics of PBF process for different laser powers.

| Case | Laser Power, P (µW) | Melt Front Heating Rate (K/ps) | Solidifying Front Cooling Rate (K/ps) | Maximum Melt-Pool Temperature, $T_{max}$ (K) |
|---|---|---|---|---|
| A | 100 | 7.86 | 4.83 | 1980 |
| K | 140 | 8.54 | 5.64 | 2326 |
| L | 180 | 9.23 | 7.06 | 2536 |
| M | 220 | 9.04 | 8.61 | 2834 |

| | **List of Figure Captions** |
|---|---|
| **Figure 1** | Illustration of MD simulation model at the beginning and end of laser melting process. The orange cylinder represents the laser, while the white dotted box represents the region from which a representative block is cut away to perform the uniaxial tensile and compression test along the Y direction. |
| **Figure 2** | Visualization of the direction of laser movement for multi-pass unidirectional and reversing cases. |
| **Figure 3** | Sectional view of the common neighbor analysis (CNA) during a single laser pass at a speed of 0.4 Å/ps portraying the melt as an amorphous region accompanied by a melting and a solidifying front. The solidified portion left behind by the laser shows the formation of several vacancies and stacking faults. |
| **Figure 4** | Common neighbor analysis (CNA) of the post-cooled structures and dislocation analysis (DLA) of the cut-off representative blocks acquired from various single-pass laser scan speeds. For clarity, only the atoms in HCP stacking faults and dislocation lines are shown in the DLA of the blocks. The other atoms depicted within the blocks signify the distribution of vacancies inside them. |
| **Figure 5** | (a) Temperature profile of PBF process, (b) tensile, and (c) compressive stress-strain relationships of the cut-off representative blocks acquired from various single-pass laser scan speeds. |
| **Figure 6** | Common neighbor analysis (CNA) of sectional view and dislocation analysis (DLA) of the cut-off representative blocks obtained from the various unidirectional laser passes at a speed of 0.4 Å/ps. Multiple laser passes over the target region annihilates certain unfixed dislocation lines and promotes the migration of free Shockley partials and HCP stacking faults away from the heat source. |
| **Figure 7** | (a) Temperature profile of PBF process, (b) tensile and (c) compressive stress-strain relationships of the cut-off representative blocks acquired from the various unidirectional laser passes at a speed of 0.4 Å/ps. |
| **Figure 8** | (a) Tensile and (b) compressive stress-strain relationships of the cut-off representative blocks acquired from various reversing laser passes at a speed of 0.4 Å/ps. |
| **Figure 9** | Variation in dislocation lengths and HCP stacking faults within the blocks obtained using single laser pass at (a) S = 0.4 Å/ps and (b) S = 0.2 Å/ps, as well as from (c) N = 2 and (d) N = 3 at 0.4 Å/ps, during uniaxial tensile testing. |
| **Figure 10** | Visualization of the internal structures of the blocks obtained from single and double laser pass at a speed of 0.4 Å/ps under tensile straining. Twin boundaries (TB) seem to form while straining N = 2 block past UTS point. |
| **Figure 11** | Variation in dislocation lengths and HCP stacking faults within the blocks obtained using single laser pass at (a) S = 0.4 Å/ps and (b) S = 0.2 Å/ps, as well as from (c) N = 2 and (d) N = 3 at 0.4 Å/ps, during uniaxial compression testing. |
| **Figure 12** | Visualization of the internal structures of the blocks obtained from single and double laser pass at a speed of 0.4 Å/ps under compressive straining. Twin boundaries (TB) seem to form while straining N = 2 block past UCS point. |
| **Figure 13** | Common neighbor analysis (CNA) of sectional view and dislocation analysis (DLA) of the cut-off representative blocks obtained from various laser power at a scan speed of 0.4 Å/ps. The melt pool forms a semi-spherical region whose diameter tends to increase with the laser power. |
| **Figure 14** | (a) Temperature profile of PBF process, (b) tensile, and (c) compressive stress-strain relationships of the cut-off representative blocks acquired from various laser powers at a speed of 0.4 Å/ps. |
| **Figure 15** | Variation in dislocation lengths and HCP stacking faults within the blocks obtained using laser powers of (a) P = 140 µW and (b) P = 220 µW during uniaxial tensile testing, as well as during uniaxial compression testing at (c) P = 140 µW and (d) P = 220 µW laser powers. |

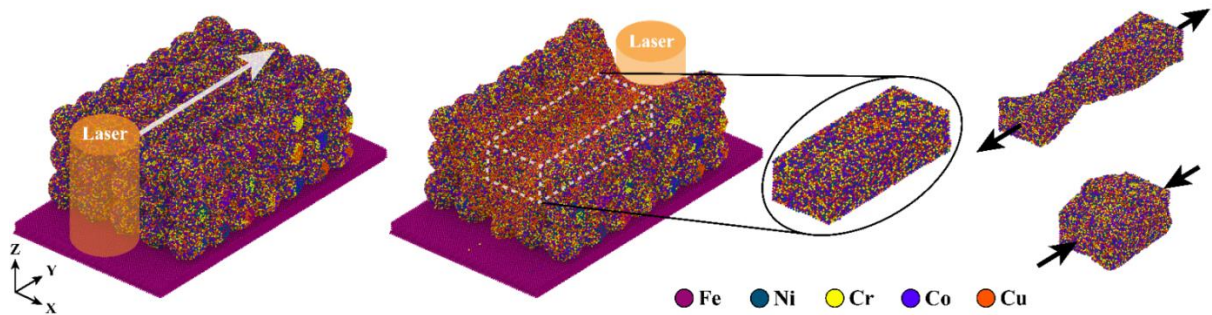

**Figure 1.** Illustration of MD simulation model at the beginning and end of laser melting process. The orange cylinder represents the laser, while the white dotted box represents the region from which a representative block is cut away to perform the uniaxial tensile and compression test along the Y direction.

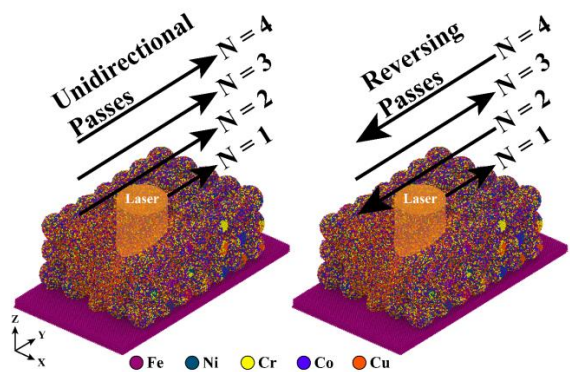

**Figure 2.** Visualization of the direction of laser movement for multi-pass unidirectional and reversing cases.

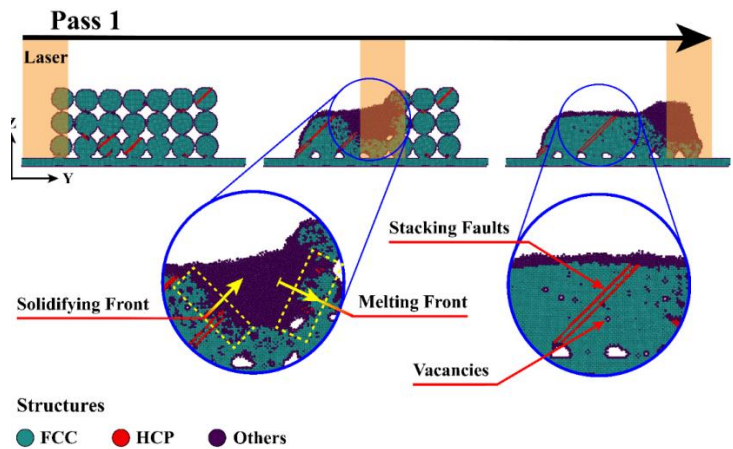

**Figure 3.** Sectional view of the common neighbor analysis (CNA) during a single laser pass at a speed of 0.4 Å/ps portraying the melt as an amorphous region accompanied by a melting and a solidifying front. The solidified portion left behind by the laser shows the formation of several vacancies and stacking faults.

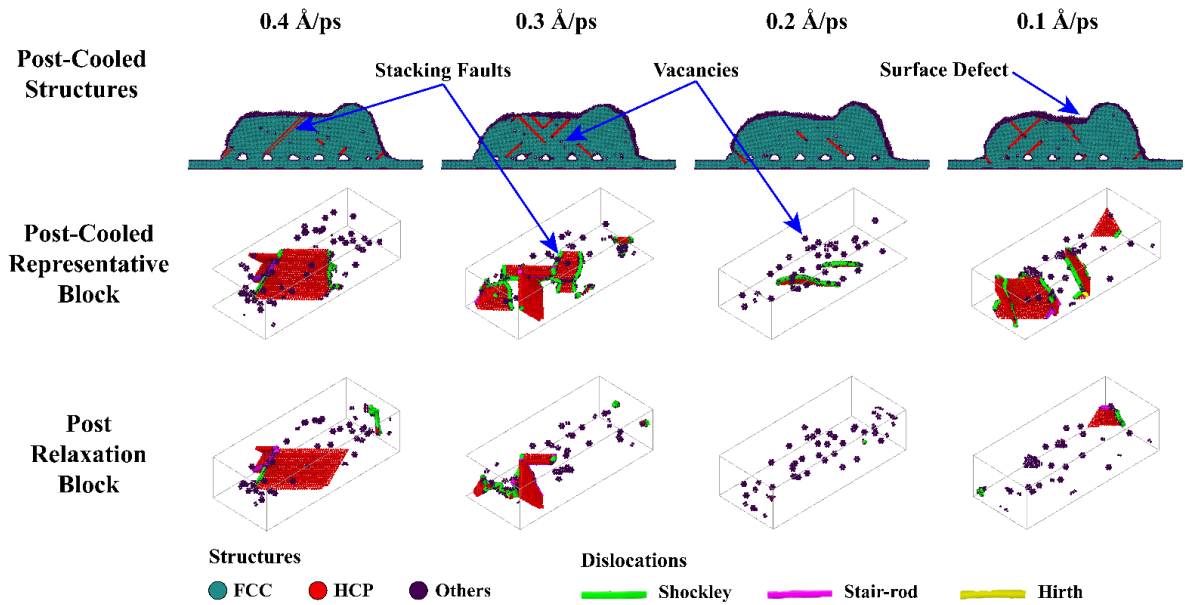

**Figure 4.** Common neighbor analysis (CNA) of the post-cooled structures and dislocation analysis (DLA) of the cut-off representative blocks acquired from various single-pass laser scan speeds. For clarity, only the atoms in HCP stacking faults and dislocation lines are shown in the DLA of the blocks. The other atoms depicted within the blocks signify the distribution of vacancies inside them.

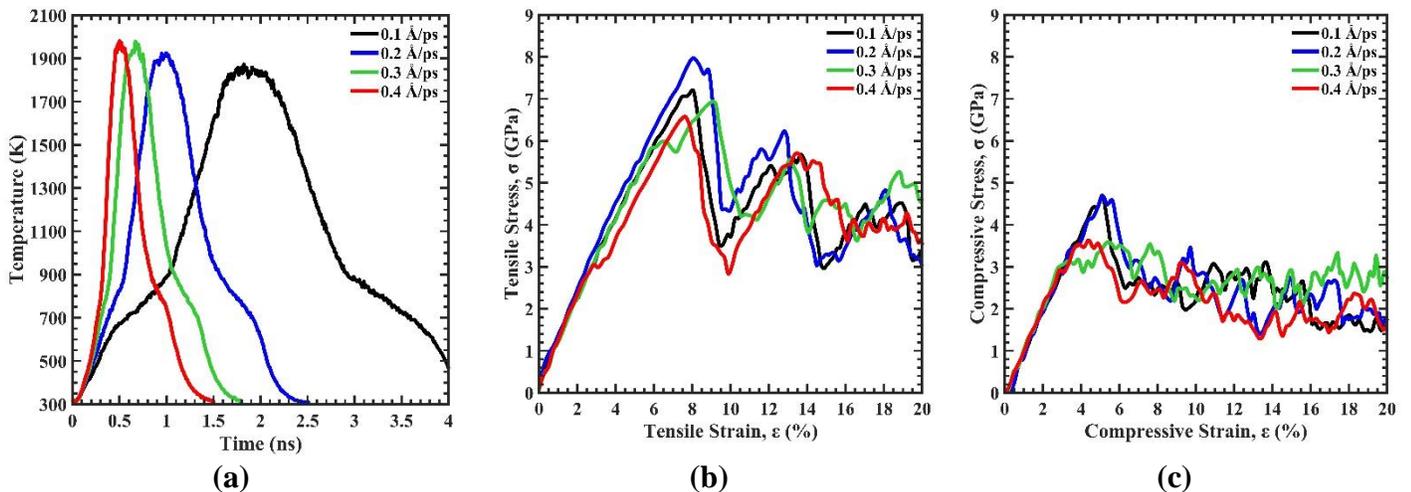

**Figure 5.** (a) Temperature profile of PBF process, (b) tensile, and (c) compressive stress-strain relationships of the cut-off representative blocks acquired from various single-pass laser scan speeds.

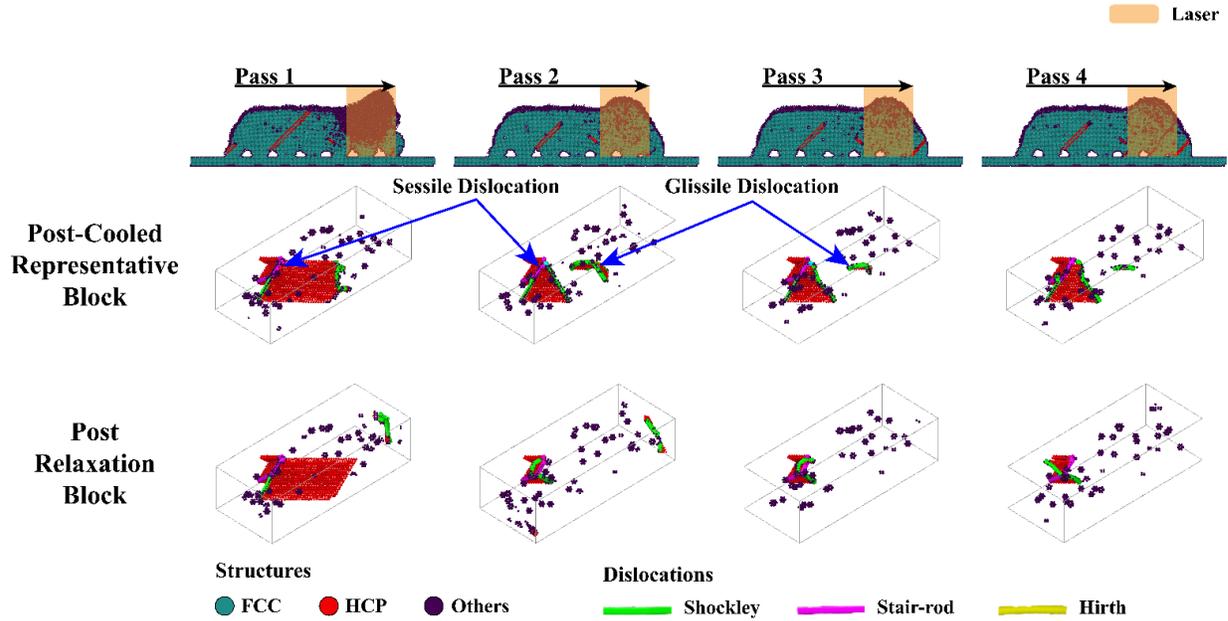

**Figure 6.** Common neighbor analysis (CNA) of sectional view and dislocation analysis (DLA) of the cut-off representative blocks obtained from the various unidirectional laser passes at a speed of 0.4 Å/ps. Multiple laser passes over the target region annihilates certain unfixed dislocation lines and promotes the migration of free Shockley partials and HCP stacking faults away from the heat source.

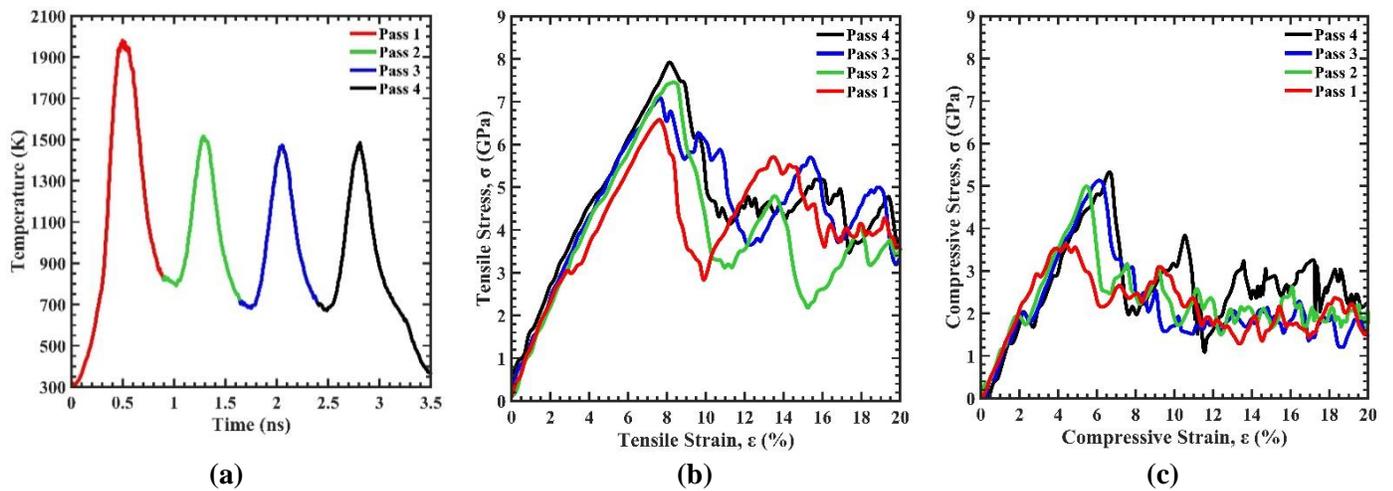

**Figure 7.** (a) Temperature profile of PBF process, (b) tensile and (c) compressive stress-strain relationships of the cut-off representative blocks acquired from the various unidirectional laser passes at a speed of 0.4 Å/ps.

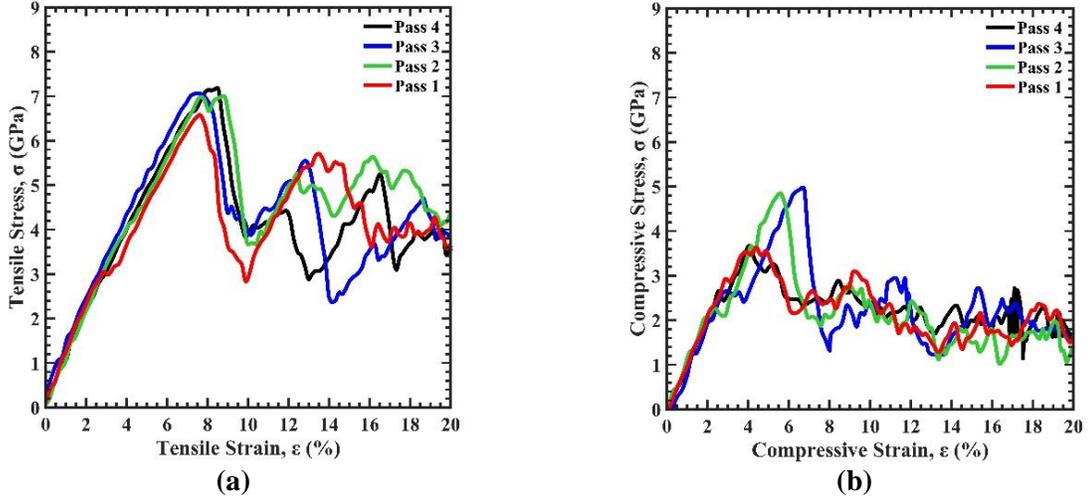

**Figure 8.** (a) Tensile and (b) compressive stress-strain relationships of the cut-off representative blocks acquired from various reversing laser passes at a speed of 0.4 Å/ps.

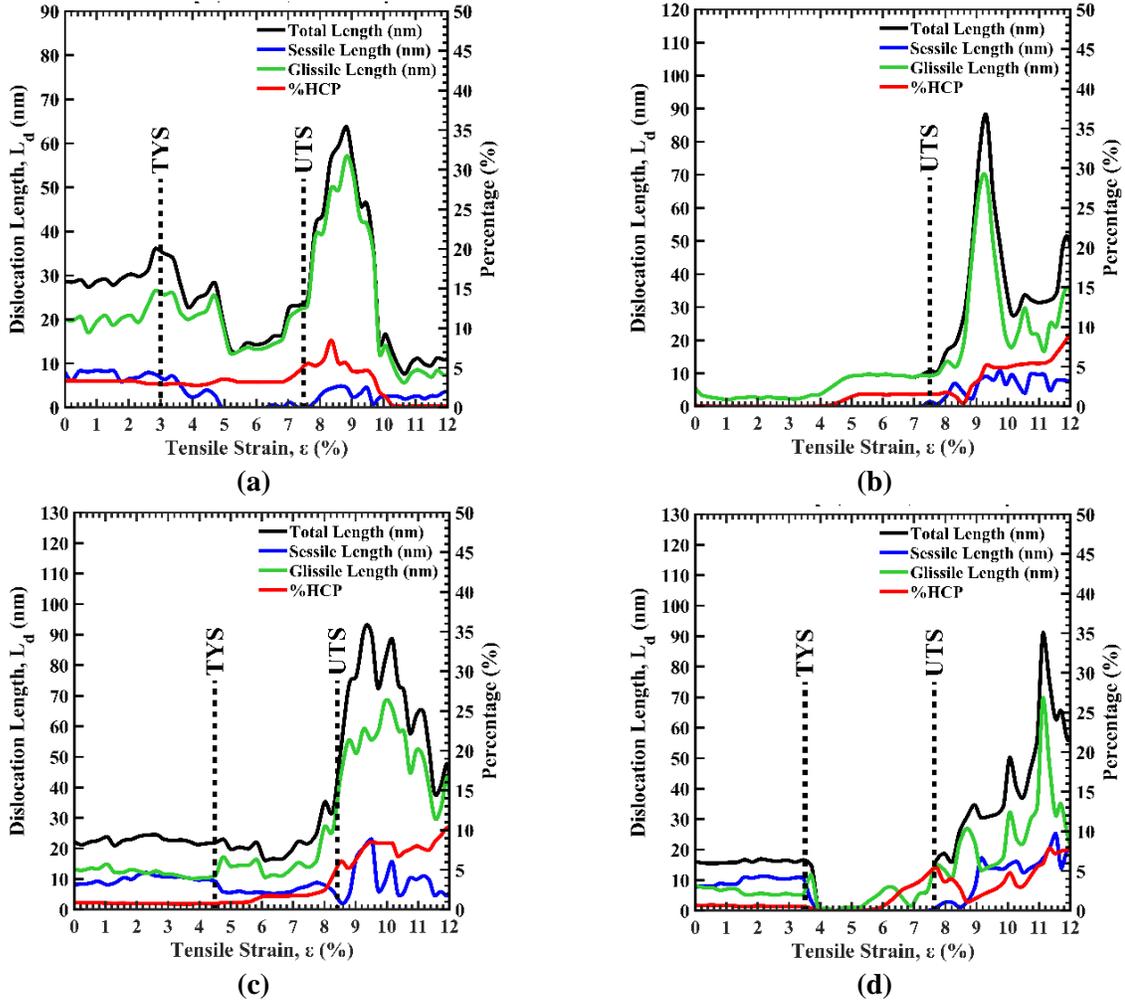

**Figure 9.** Variation in dislocation lengths and HCP stacking faults within the test blocks obtained for P = 100 μW using single laser pass at (a) S = 0.4 Å/ps and (b) S = 0.2 Å/ps, as well as from (c) N = 2 and (d) N = 3 at 0.4 Å/ps, during uniaxial tensile testing.

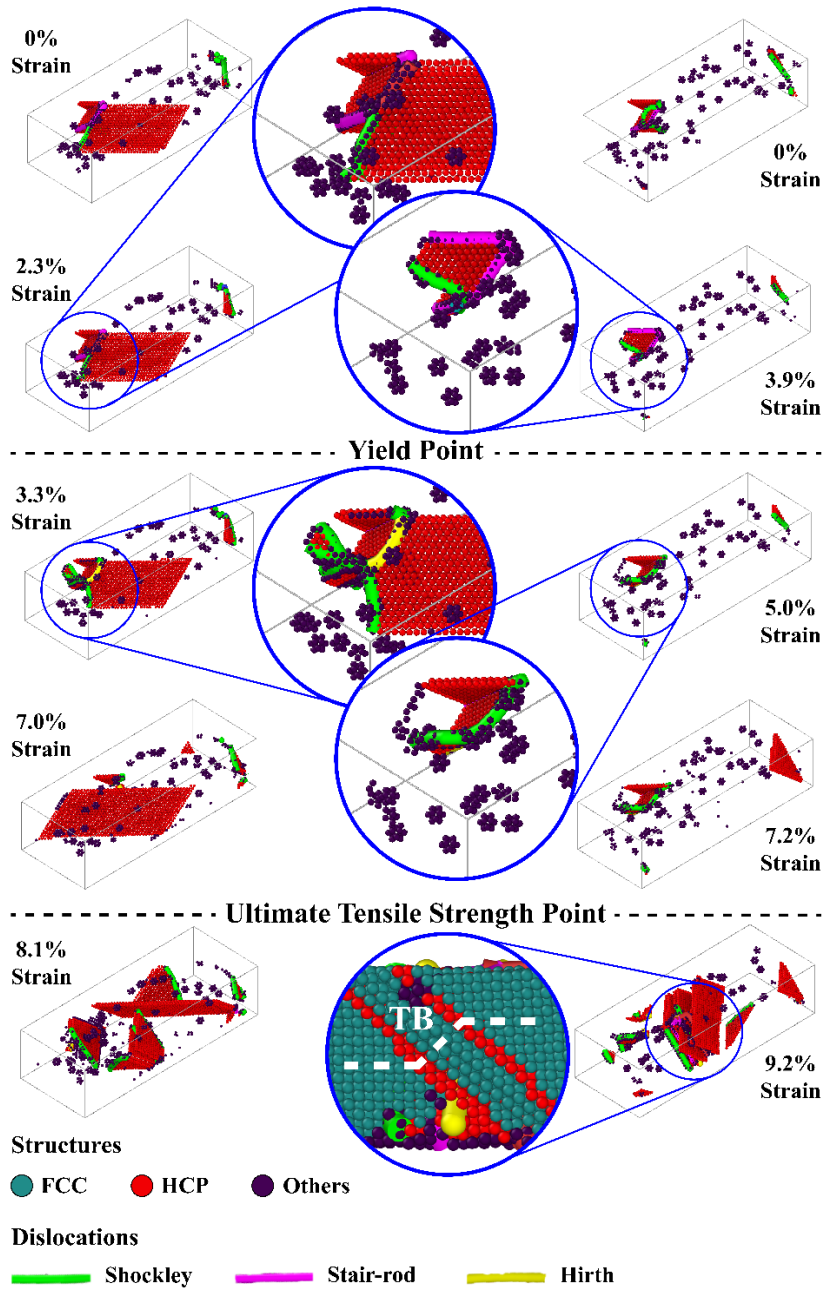

**Figure 10.** Visualization of the internal structures of the blocks obtained from single and double laser pass at a speed of 0.4 Å/ps under tensile straining. Twin boundaries (TB) seem to form while straining N = 2 block past UTS point.

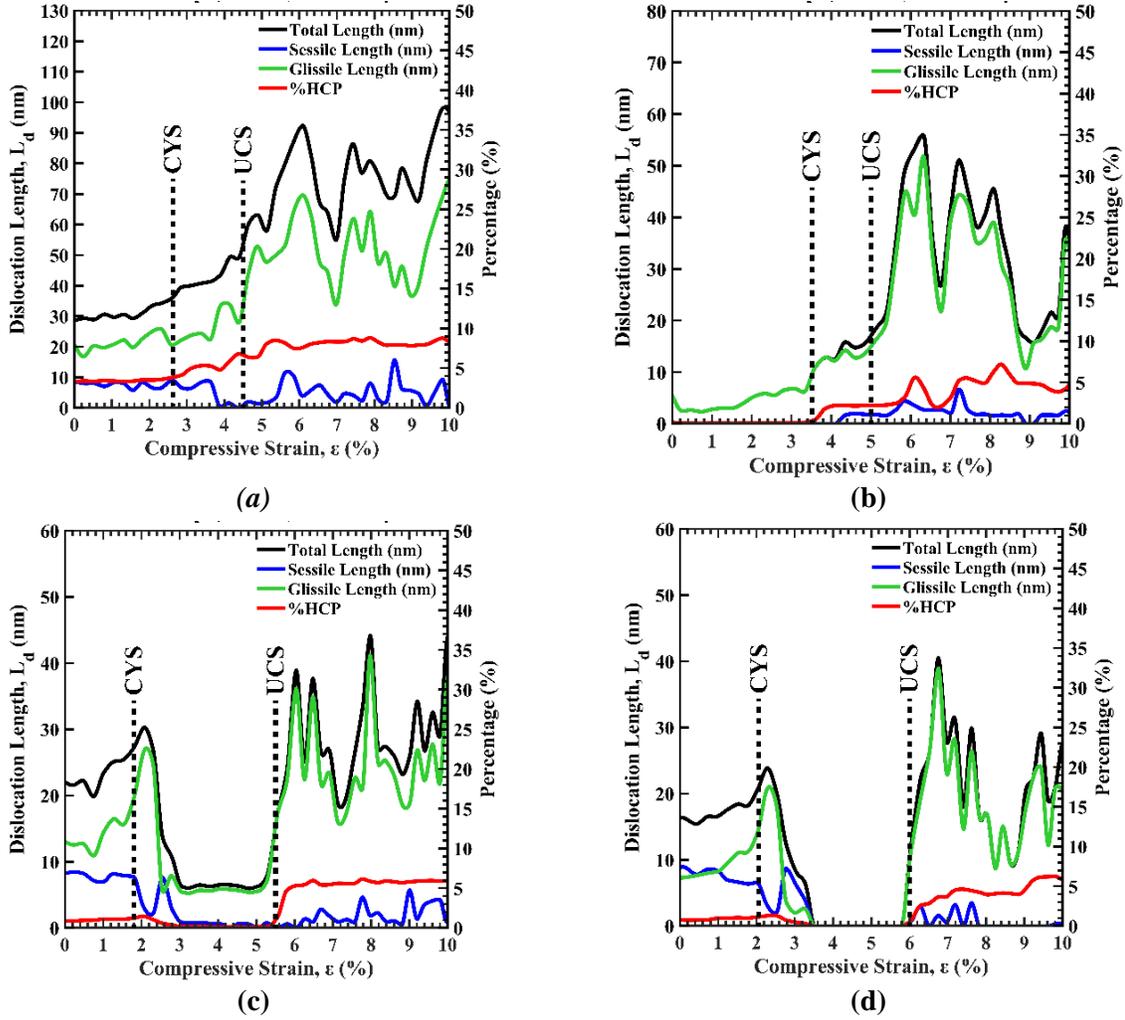

**Figure 11.** Variation in dislocation lengths and HCP stacking faults within the test block obtained for P = 100 μW using single laser pass at (a) S = 0.4 Å/ps and (b) S = 0.2 Å/ps, as well as from (c) N = 2 and (d) N = 3 at 0.4 Å/ps, during uniaxial compression testing.

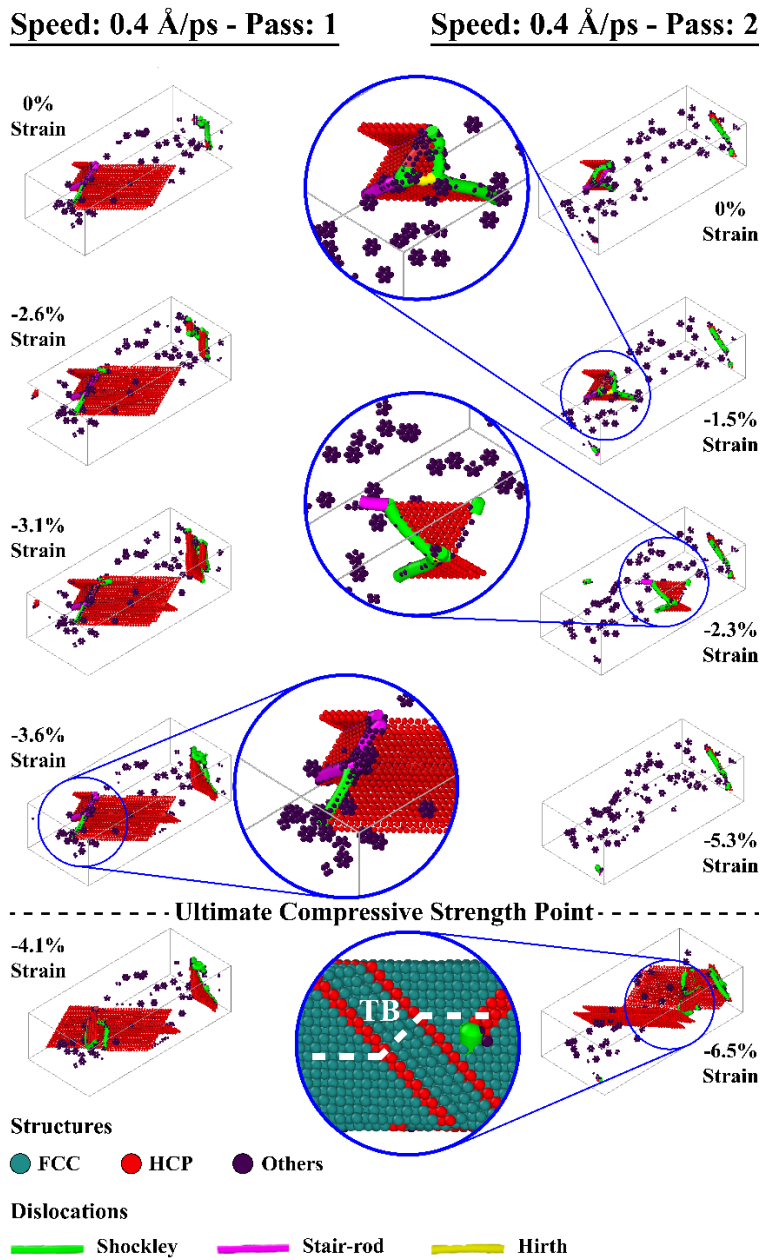

**Figure 12.** Visualization of the internal structures of the blocks obtained from single and double laser pass at a speed of 0.4 Å/ps under compressive straining. Twin boundaries (TB) seem to form while straining N = 2 block past UCS point.

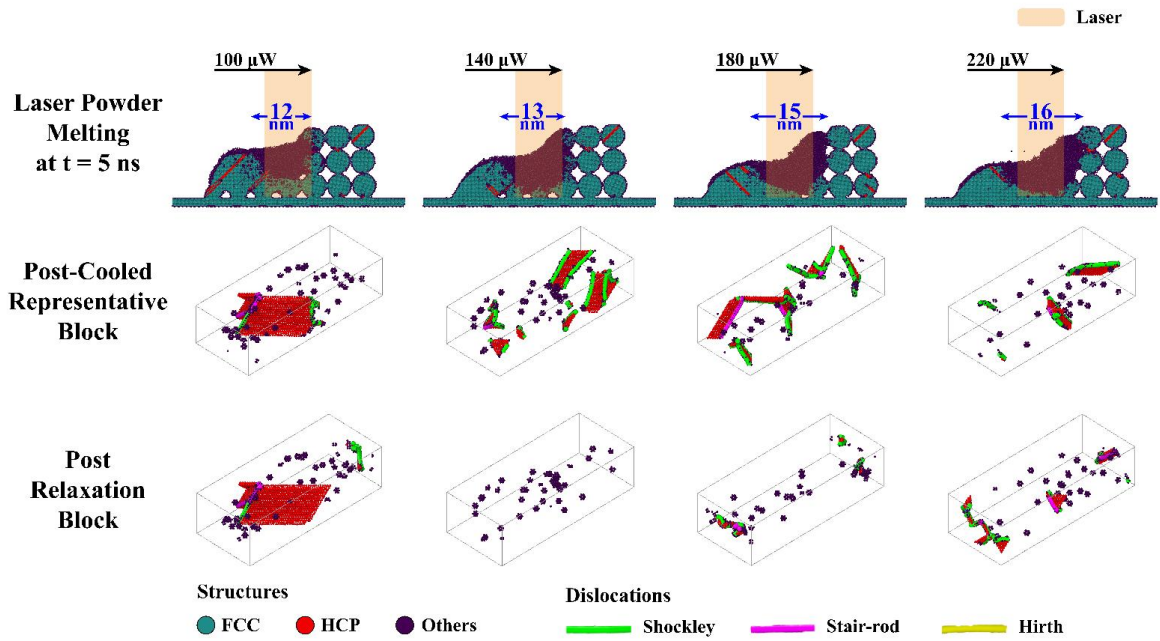

**Figure 13.** Common neighbor analysis (CNA) of sectional view and dislocation analysis (DLA) of the cut-off representative blocks obtained from various laser power at a scan speed of 0.4 Å/ps. The melt pool forms a semi-spherical region whose diameter tends to increase with the laser power.

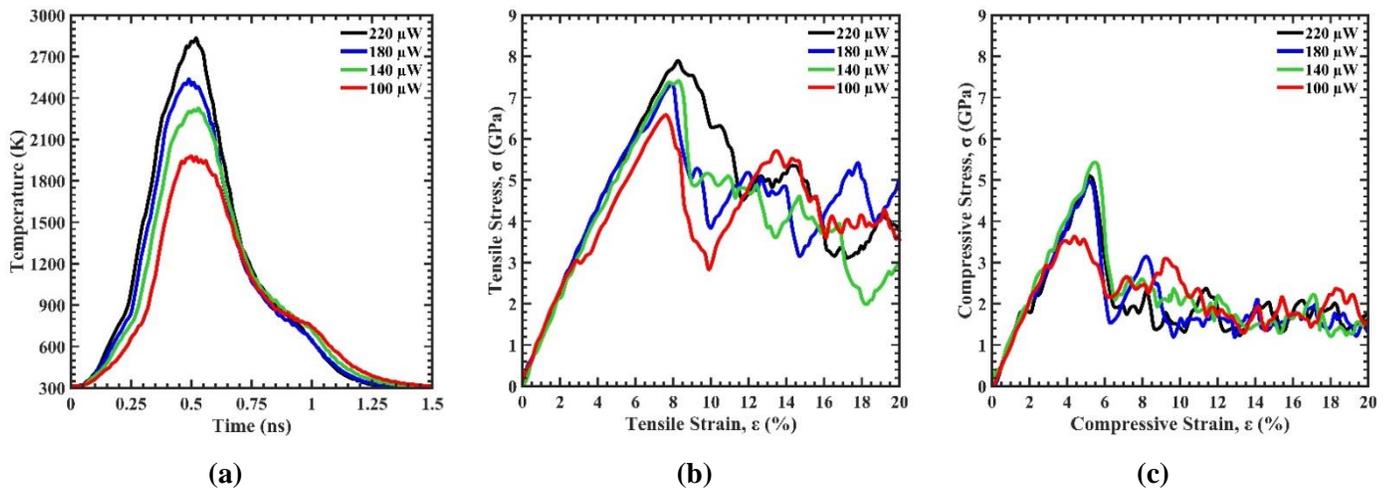

**Figure 14.** (a) Temperature profile of PBF process, (b) tensile, and (c) compressive stress-strain relationships of the cut-off representative blocks acquired from various laser powers at a speed of 0.4 Å/ps.

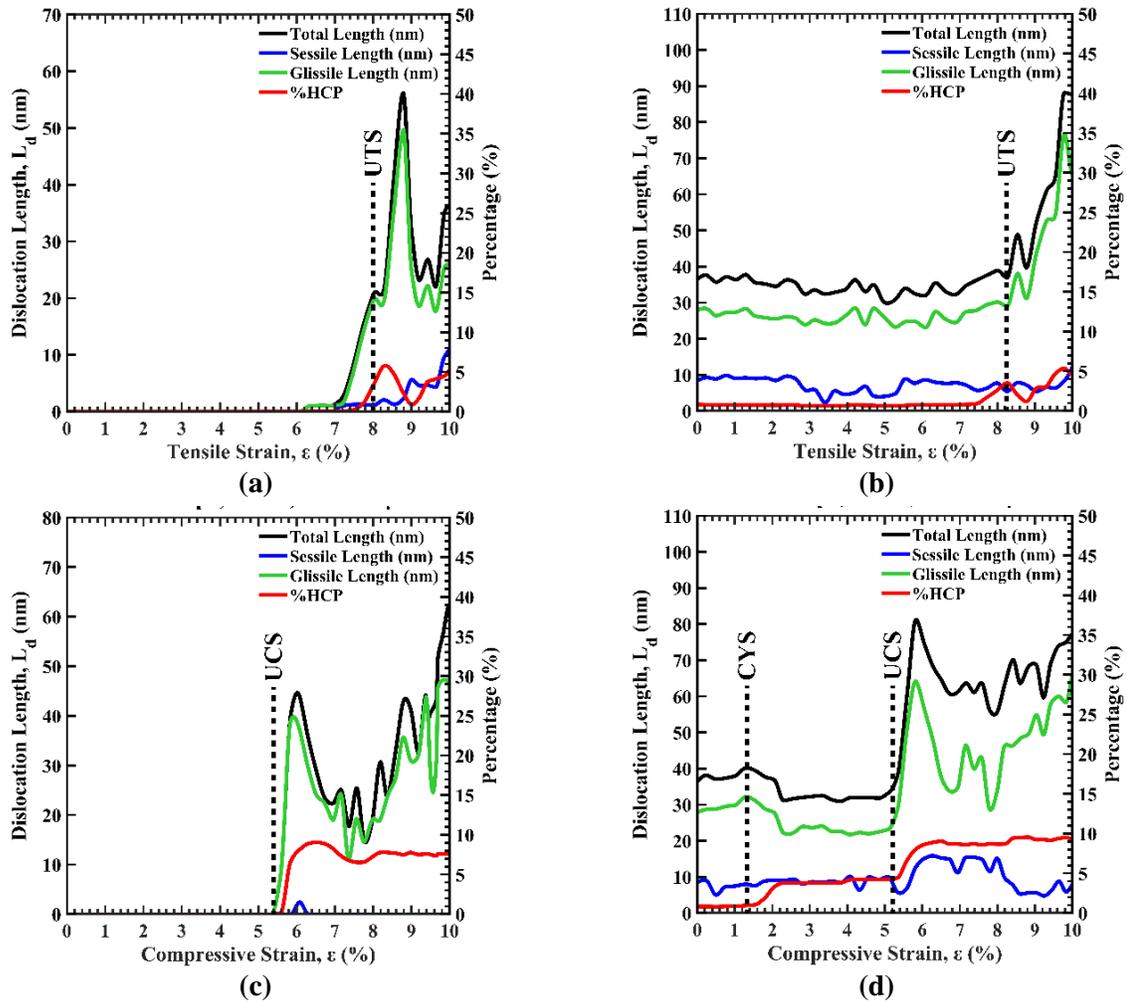

**Figure 15.** Variation in dislocation lengths and HCP stacking faults within the test block obtained for single pass, laser speed of S = 0.4 Å/ps with laser power of (a) P = 140 µW and (b) P = 220 µW during uniaxial tensile testing, as well as during uniaxial compression testing at (c) P = 140 µW and (d) P = 220 µW laser powers.